\documentclass[12pt]{article}
\usepackage{amsfonts,amsmath,amssymb,epsf,cite}
\usepackage{graphicx,color}
\usepackage[usenames,dvipsnames]{pstricks}
 \def\ep{{\epsilon}}

 \def\frac#1#2{{#1\over #2}}

 \def\s{\sqrt}

 \def\al{\alpha'}
 \def\de{\partial}

 \def\f {\frac}
 \def\ti{\tilde}
 \def\ap{\alpha}

 \def\ddd{\cdot\cdot\cdot}
 \def\no{\nonumber \\}

 \def\la{\langle}
 \def\lb{\rangle}
 \def\ep{\epsilon}

 \def\vp{\varphi}

 \def\ep{{\epsilon}}

 \def\frac#1#2{{#1\over #2}}

 \def\s{\sqrt}

\def\be{\begin{equation}}
\def\ee{\end{equation}}
\def\ba{\begin{eqnarray}}
\def\ea{\end{eqnarray}}

\begin{document}
\begin{titlepage}
\thispagestyle{empty}
\begin{flushright}
IPMU10-0142
\end{flushright}

\bigskip

\begin{center}
\noindent{\large \textbf{Measuring Black Hole Formations by Entanglement
Entropy\hspace{5cm}
via Coarse-graining}}\\
\vspace{2cm} \noindent{ Tadashi
Takayanagi\footnote{e-mail:tadashi.takayanagi@ipmu.jp} and Tomonori
Ugajin\footnote{e-mail:tomoki.ugajin@ipmu.jp}}

\vspace{1cm}
  {\it
Institute for the Physics and Mathematics of the Universe (IPMU), \\
 University of Tokyo, Kashiwa, Chiba 277-8582, Japan\\
 }
\end{center}
\vspace{1cm}
\begin{abstract}
We argue that the entanglement entropy offers us a useful coarse-grained entropy in time-dependent
AdS/CFT. We show that the total von-Neumann entropy
remains vanishing even when a black hole is created in a gravity dual,
being consistent with the fact that its corresponding CFT is described by a time-dependent pure state.
We analytically calculate the time evolution of entanglement entropy for a free Dirac fermion on a circle
following a quantum quench. This is interpreted as a toy holographic dual of black hole creations and
annihilations. It is manifestly free from the black hole information problem.

\end{abstract}
\end{titlepage}
\newpage

\section{Introduction}

The concept of holography, especially the AdS/CFT correspondence,
has now been the most powerful tool to analyze quantum aspects of
gravity \cite{tH,Su,Maldacena}. Since there have been no
satisfactory and non-perturbative formulations of string theory, the
direct calculations of quantum gravity effects often turn out to be
quite difficult. However, the AdS/CFT equivalently transforms the
problems into those in non-gravitational theories, typically large
$N$ gauge theories \cite{Maldacena}. Since the latter ones are
non-perturbatively well defined, we can in principle understand all
quantum gravity effects at least in AdS spaces.

In spite of enormous related developments, the AdS/CFT in
time-dependent backgrounds have not been well understood at present.
One of purposes of this paper is to understand better somewhat
puzzling aspects of entropy in
 time-dependent backgrounds, which is also closely related to the black hole
 information problem \cite{Ha}. If we suddenly excite a CFT, it will eventually
 thermalize after a certain time. In AdS/CFT, this is dual to a black hole
 formation in AdS spaces \cite{ChYa,Bhattacharyya:2009uu}. In a probe D-brane
 approximation, a thermalization is dual to a black hole formation in the
 induced metric on the brane \cite{DNT}.

In such a process, it is clear that the density matrix of the system
is described by a time-dependent pure state in the CFT side,
assuming that we start with a pure state before the thermalization.
Therefore, the von-Neumann entropy is always vanishing. On the other
hand, in the AdS gravity, since the black hole will be formed and
the spacetime at late time is well approximated by the AdS
Schwarzschild black hole, the entropy seems non-vanishing. For
example, it is clear that the entropy which appears in the fluid
description gets non-vanishing \cite{Bha}. This might suggest an
evolution of a pure state into a mixed state, which contradicts with
the principle of quantum mechanics. To resolve this issue, we need
to distinguish the microscopic von-Neumann entropy from a
coarse-grained entropy. Only the latter can get non-vanishing in
this thermalization process. We will argue that the entanglement
entropy \cite{BoSr,CaCaR,CaHuR} offers us a convenient
coarse-grained entropy in holographic setups. At the same time, we
will show that the von-Neumann entropy of the total system is always
vanishing in the black hole formation process employing the
holographic calculation of the entanglement entropy
\cite{RT,HRT,EEReview}.

From the CFT viewpoint, the black hole formation process is dual to
a problem of non-equilibrium dynamics and therefore is quite
complicated to analyze explicitly in general. Fortunately, recently
there have been many progresses in understanding of properties of
quantum field theories after thermalizations induced by
instantaneous excitations called quantum quenches
\cite{CCa,CCb,Inte,CCc,SCa,CCd,SCb}. We will consider a two
dimensional CFT defined by a free Dirac fermion on a circle and
calculate the time-evolution of the entanglement entropy and
correlation functions after a quantum quench, taking into account
the finite size effect. Indeed, the behavior of the entanglement
entropy we obtained implies that its gravity dual is supposed to be
successive creations and annihilations of black holes in AdS spaces,
though a free field theory in general corresponds to an extremely
stringy region of the gravity side. This solvable example offers us
a CFT description of a black hole formation and decay, which are
manifestly free from the black hole information problem, because the
system is clearly described by a pure state. Notice that in this
problem, the standard thermal entropy is not useful as it is
vanishing and the entanglement entropy is an alternative important
quantity which can directly probe black holes. 
 
 We would also like to mention that quite recently, remarkable
numerical calculations of time evolutions of entanglement entropy
have been done holographically in \cite{AAL,AC}. When we essentially finished 
calculations for this paper, 
we found the paper \cite{AAL},
which has a partial but important overlap with ours
about the argument which shows unitary evolutions of CFTs dual to
the black hole formations by using the holographic entanglement
entropy. A similar issue has also been
discussed independently in the appendix of the paper \cite{AC}.

The paper is organized as follows: In section two, we will explain a
puzzle on the entropy in time-dependent backgrounds and resolve it
by employing the entanglement entropy as a coarse-grained entropy.
In section three, we will calculate the time evolving entanglement
entropy in two dimensional CFT of a free Dirac fermion with finite
size effect. We will interpret our results from the gravity dual
viewpoint and relate them to the black hole creations and
annihilations. In section four, we will study one and two point
functions in our time-dependent CFT and confirm that they agree with
our holographic interpretation. In section five, we will summarize
our conclusion. In the appendix A, we present detailed calculations
of two point functions in a free Dirac fermion theory on a cylinder
using the boundary state formalism.

\section{Holographic Entanglement Entropy as Coarse-grained Entropy}

\subsection{An Entropy Puzzle about Time-dependent Holography}

Consider holography in time-dependent backgrounds, especially the ones in AdS/CFT.
We are particularly interested in a situation where a black hole horizon is created by a
gravitational collapse. Such a time-dependent process has been considered to be dual to
thermalizations and indeed we can construct explicit examples in AdS/CFT setups
\cite{ChYa,Bhattacharyya:2009uu}, where non-normalizable perturbations lead to a creation of
hot plasma fluid in the dual Yang-Mills theory. Similar phenomena also occur in the probe D-brane setups
regarding the thermalization as a formation of horizon in the induced metric \cite{DNT}.

An important characteristic physical quantity in time-dependent systems will be the entropy.
Since the black hole is created after a certain time in our
gravity background, it is quite natural that the entropy which is holographically calculated, gets
non-vanishing at late time. However, in the dual CFT, the time evolution is described by a time-dependent
external force in quantum many-body systems and thus the density matrix $\rho_{tot}(t)$ of the total system will evolve in a unitary way $\rho(t)=U(t,t_0)\rho(t_0)U(t,t_0)^{-1}$. Since the entropy of the total system is defined by the von-Neumann entropy
\be
S_{tot}(t)=-\mbox{Tr}[\rho(t)\log\rho(t)]=S_{tot}(t_0),
\ee
which is clearly time-independent. Therefore if we start with a pure state, then the entropy is always
vanishing at any time, even if the black hole horizon appears in the dual geometry.
This observation seems to lead to a contradiction between the gravity and CFT at first sight.

Another confusing point is that
there is no known unambiguous definition of entropy $S_{tot}$ in the time-dependent gravity duals.
This is partly because there are two different notions of horizons: event horizon and apparent horizon. Currently, there are several evidences that the latter will be more appropriate to define entropy \cite{HRT,FHRR}. However, still we need to specify the choice of time slices to calculate the
apparent horizon and thus there are infinitely many different definitions.
Remember that in static spacetimes, the time slice is uniquely chosen and there is no ambiguity.

A closely related issue in the dual CFT side is that we can define a non-vanishing entropy
even for a pure state if we perform a coarse-graining of the given total system. If the total
system consists of a gas and a heat bath, then the entropy of gap is obtained by tracing out the
heat bath and by the required coarse-graining due the fact that our observables
are quite restricted compared with the microscopic degrees of freedom; see e.g. \cite{Brown}
for relevant quantum field theoretic formulations. Again this
leads to infinitely many different definitions of (coarse-grained) entropy.
Below we would like to resolve the above puzzle by paying attention to this coarse-graining
procedure.

The resolution of this puzzle is also closely related to that of the black hole information loss problem.
This problem occurs since a massive object described by a pure state collapses into a black hole and
finally ends up with a thermal gas generated by Hawking radiations, which looks like a mixed state.
If this is true, a pure system has to evolve into a mixed state after enough time and clearly
contradicts with the principle of quantum mechanics.

\subsection{Entanglement Entropy as Coarse-grained Entropy}

Typically the coarse-graining is done by cutting out higher energy modes or higher order
multi-particle interactions. However, this does not seem to allow straightforward calculations
in AdS/CFT setup. Therefore, here we would like to perform a coarse-graining by cutting out a
spacial part of the total system. In this case, the obtained entropy coincides with the quantity
called entanglement entropy.

The entanglement entropy $S_A(t)$ is defined for each subsystem $A(t)$ of the total system. In quantum
field theories, $A(t)$ is specified by diving the total space at a fixed time $t$ into two parts
named as $A(t)$ and $B(t)$. Thus the number of choices of $A$ are obviously infinite.
The total Hilbert space
$H_{tot}$ now becomes a direct product $H_{tot}=H_{A}\otimes H_{B}$, where the subsystem $B$ is
the complement of $A$. The reduced density matrix for $A$ at any time $t$ is given by
\be
\rho_A(t)=\mbox{Tr}_{H_B}\rho_{tot}(t).
\ee
Finally we define the entanglement entropy as follows
\be
S_A(t)=-\mbox{Tr}[\rho_A(t)\log\rho_A(t)].
\ee
This quantity is in general non-vanishing and a non-trivial function of time even for pure states.

We can calculate $S_A(t)$ via AdS/CFT by applying the holographic
formula proposed in \cite{HRT}, which generalizes the holographic
formula in static spacetime found in \cite{RT}. It is simply given
by \be S_A(t)=\f{\mbox{Area}\left(\gamma_A(t)\right)}{4G_N}, \ee
where $G_N$ is the Newton constant of the AdS space. The surface
$\gamma_A(t)$ is a codimension two surface in the AdS space which is
defined by the extremal surface whose boundary coincides with the
boundary of $A(t)$. We require that $\gamma_A(t)$ is homotopic to
$A(t)$. If there are more than one such extremal surfaces, we pick
up the one with the lowest area, as is required by the strong
subadditivity of entanglement entropy \cite{HeTa}. Note that here we
need to directly deal with Lorentzian spacetime without its Euclidean
counterpart because it is time-dependent. Indeed, in \cite{HRT}, it
was confirmed that such extremal surfaces are well-defined in
important examples. Recently, this formula is applied to the AdS$_3$
Vaidya background and the conformal field theory results in
\cite{CCa} have been remarkably reproduced in the work \cite{AAL}.
Moreover, this calculation has been extended to the higher
dimensional cases in \cite{AC}.

For our purpose, it is convenient to study a $d+1$ dimensional time-dependent background which is
asymptotically global AdS$_{d+1}$ space, which includes a finite size effect because
its boundary is $R\times S^{d-1}$. We assume that a black hole is produced at a certain
time via a gravitational collapse and this process lasts only for a finite time. Our argument
below is heuristic and does not depend on the detailed form of the time-dependent solution.

\begin{figure}
\begin{center}
\includegraphics[height=7cm,clip]{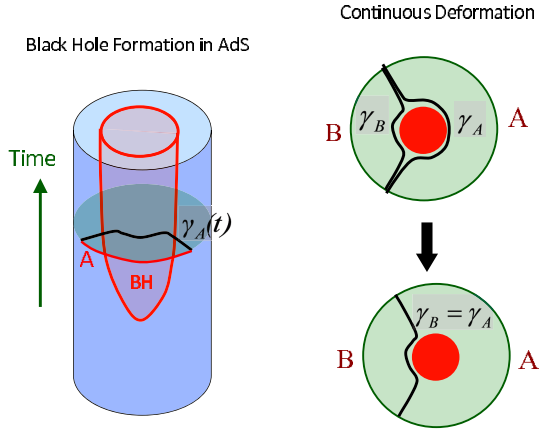}
\end{center}
\caption{The minimal surface $\gamma_A$ for the holographic calculation of entanglement entropy in
the AdS black formation (left). Though for an eternal AdS black hole we have $\gamma_A\neq \gamma_B$
(upper right), in our case of black hole formation, we actually find that $\gamma_A=\gamma_B$
(lower right). This is because
the horizon vanishes at early time.}\label{BHf}
\end{figure}

We calculate $S_A$ at a time $t$ when the black hole formation has
been ended and when the spacetime is well approximated by a static
AdS black hole. First we assume the length size of $A$ (denoted by
$|A|$ below) is smaller than the inverse temperature $\beta_{BH}$ of
the AdS black hole temperature. In this case, the extremal surface
$\gamma_A(t)$ is localized near the boundary and surrounds the
region $A(t)$ \cite{RT}. $S_A$ essentially consists of the area law
divergent piece \be S_A\sim \f{\mbox{Area}(\de A)}{a^{d-2}}+....,
\label{area} \ee which are independent of the temperature. Notice
that the condition $|A|<<|B|$ assumed here physically means that the
system is heavily coarse-grained. As $|A|$ exceeds $\beta_{BH}$,
$S_A$ gains a finite and extensive contribution, which is
essentially the thermal entropy for the subsystem $A$, in addition
to the divergent part $S_{div}$ given by (\ref{area}). This thermal
contribution comes from the part of the surface $\gamma_A(t)$ which
is wrapped on a part of the apparent horizon of the AdS black hole
and thus is proportional to $|A|$ as in Fig.\ref{BHf}; see
\cite{HRT,HR,AAL,AC} for explicit confirmations.  An important fact
is that though the apparent horizon requires the choice of the time
slice, our extremal surface $\gamma_A(t)$ is uniquely determined by
the boundary condition.

When this reaches the point $|A|=|B|$, the situation begins to change importantly.
If we naively speculate the surface $\gamma_A(t)$ in a continuous way until $|A|$ exceed $|B|$,
we may think that it will wrap on a more than half of the apparent horizon
as in the right-up figure of Fig.\ref{BHf}. However,
since the horizon disappears at an earlier time, we can smoothly deform this surface into the one
which wraps the opposite part of the horizon. Therefore the result of
$S_A$ as a function of $|A|$ becomes symmetric with respect to the point $|A|=|B|$ as in
the right-down figure of Fig.\ref{BHf}. In other words, we can
conclude that in this time-dependent background the entanglement entropy satisfies
\be
S_A=S_B,
\ee
which is indeed known to be true when the total system is a pure state and thus agrees with our holographic
setup. This sudden decreasing of the entanglement entropy occurs because the subsystem $B$ which we are
tracing out gets smaller than the half of the total system and the information lost by the coarse-graining starts recovering\footnote{A similar
behavior has been known in the analysis of quantum information in evaporating black holes \cite{Page}.
This has been considered to be crucial to resolve the information paradox of quantum black holes
because the information may be recovered after more than half of a black hole has been evaporated.}.

In the end, we can holographically calculate the von-Neumann entropy for the total system
as follows
\be
S_{tot}=\lim_{|B|\to 0} \left(S_{A}-S_{B}\right)=0. \label{entroth}
\ee
Thus, in the gravity side, we confirmed that the total entropy is always vanishing in spite of the
black hole formation. As is clear from the previous argument, the essential point is that the
horizon is time-dependent and gets vanishing at earlier time and the minimal area principle (among extremal surfaces) prefers the minimum i.e. $S_{tot}=0$ at any time.
In summary, in the time-dependent background which describes a thermalization of a pure state, the
von-Neumann entropy of the total system is indeed vanishing also in the gravity side.
This contrasts strikingly with the setup of an eternal AdS black hole \cite{MaE} as
it is dual to a mixed state in a thermal CFT and has a non-vanishing thermal entropy, which precisely
coincides with the value $\lim_{|B|\to 0} \left(S_{A}-S_{B}\right)$ \cite{ANT}.

The non-zero entropy is obtained only after a certain coarse-graining.
Among various such entropies defined in the dual CFT, only the entanglement entropy has its clear
holographic dual in time-dependent backgrounds at present. From this viewpoint, we can define a coarse-grained effective entropy $S_{eff}$ which looks analogous to the thermal entropy by
\be
S_{eff}=2\left(S_{A}-S_{A}^{(0)}\right)\Bigl|_{|A|=|B|}. \label{effE}
\ee
We subtracted the entanglement entropy $S_A^{(0)}$ before the thermalization. This clearly includes
the divergent part $S_{div}$ given by (\ref{area}) and therefore $S_{eff}$ is finite. The reason
why we put the factor two is $A$ covers only a half of the total system. We would like to argue that
(\ref{effE}) is a definition of coarse-grained entropy which is uniquely calculable both in
gravity and CFT side of the AdS/CFT. Notice that though the calculation of entropy
from an apparent horizon requires the choice of time slice and is ambiguous, our entropy $S_{eff}$ has no
ambiguity as the extremal surface condition of $\gamma_A(t)$ determined where the surface wraps the
apparent horizon.

Finally, we would like to come back to the relation to the black hole information problem \cite{Ha}.
From the above argument, it is clear that there is no information loss in the gravitational
collapsing process because $S_{tot}=0$ is always satisfied. However, we cannot study
the most important process of the decay
due to Hawking radiations because the absence of the holographic formula of entanglement entropy
beyond the supergravity approximation. Therefore, for this purpose, it is a better idea instead
to study its CFT dual, directly. We will analytically investigate such an example in the next section.

\section{Time-dependent Entanglement Entropy in 2D CFT with finite size via Quantum Quench}

In the previous section, we studied the behavior of coarse-grained entropy by using
the holographic entanglement entropy. In this section, we would like to complete
the argument by presenting an explicit calculations with a finite size effect
in two dimensional CFT and by
showing that the results agree with the holographic ones. We would like to analyze how
the entanglement entropy evolves in time-dependent backgrounds which describe the thermalization
phenomena. At the same time, this offers us a toy gravity dual of a black hole formation and evaporation,
where there is no information loss problem and where we can solve the model exactly.

We will particularly look at a specific class of time-dependent system which is called
quantum quench, whose field theory descriptions have been
given in \cite{CCa,CCb,Inte,CCc,SCa,CCd,SCb}. The quantum quench is a phenomenon when we suddenly change
parameters of a given system such as masses and coupling constants. The evolutions of
entanglement entropy under quantum quenches have been studied in \cite{CCa,CCc,SCa} for infinitely
extended systems.

To make an exact calculation possible, we consider a free Dirac fermion theory in two dimension.
In free field theories, the thermalization via the quantum quench is not a standard one
described by the (grand-)canonical ensemble because the
different momentum modes are decoupled and do not mix \cite{Inte,CCd}. Notice also that the free CFT
corresponds to string theory with a lot of quantum corrections via the AdS/CFT.
Nevertheless, this system is enough for our purpose qualitatively as we will see later.
As far as the authors know, our result will be the first
analytical calculation of time-dependent entanglement entropy with a finite size effect. The calculation of entanglement entropy at finite temperature and with finite size has
been calculated in \cite{ANT} for a free Dirac fermion.

\subsection{How to Calculate Entanglement Entropy under Quantum Quench}

The setup of quantum quench that we have in mind is as follows. Start with a Dirac fermion theory with a mass $m$ in two dimension. Then at the time $t=0$, suddenly we turn off the
fermion mass. Then the system after this quantum quench i.e. $t>0$ will be described by a time-dependent
background in the $c=1$ 2D CFT defined by a massless Dirac fermion whose Hamiltonian is written as $H$.
If we denote the state just after the quench by $|\Psi_0\lb$, then the density matrix at a time $t>0$ is given by
\be
\rho_{tot}(t)=e^{-itH}|\Psi_0\lb \la \Psi_0| e^{itH}.
\ee
To find $|\Psi_0\lb$ we would like to follow the prescription in \cite{CCa}. We expect that in the infrared limit
$|\Psi_0\lb$ should flow into a boundary fixed point. Therefore we can approximate  $|\Psi_0\lb$ by a boundary state $|B\lb$ with the UV modes filtered out
\be
|\Psi_0\lb\simeq e^{-\ep H}|B\lb,
\ee
where $\ep$ is the parameter of UV filter and is inversely proportional to the original fermion mass
$\ep\sim m^{-1}$. If we denote the UV cut off or lattice spacing of this field theory by $a_{UV}$, we need to
take $\ep>>a_{UV}$, which is clearly satisfied in the continuum limit $a_{UV}\to 0$.

In this way we find that the reduced density matrix is eventually written as
\be
\rho_{tot}(t)=e^{-itH-\ep H}|B\lb \la B| e^{itH-\ep H}. \label{totd}
\ee

In the path-integral description \cite{CCa}, the spacetime where the Dirac fermion lives
is the cylinder geometry given by the direct product of  an interval (i.e. Euclidean time)
and a circle (i.e. space).
The length of the interval is $2\ep$ and the one of the circle is normalized into $2\pi$.
We introduce a complex coordinate $(y,\bar{y})$ for this spacetime such that
\be
y=\tau-i\sigma,\ \ \ \bar{y}=\tau+i\sigma, \ \ \
(0\leq \tau \leq 2\ep,\  0\leq \sigma \leq 2\pi). \label{cyl}
\ee
The Dirac fermion $(\psi,\bar{\psi})$ on this spacetime is decomposed into
the left-moving part $(\psi_L(y),\bar{\psi}_L(y))$
and the right-moving part $(\psi_R(\bar{y}),\bar{\psi}_R(\bar{y}))$.

We define the subsystem $A(t)$ as an interval in the Im$y$ direction at time $\ep+it$.
We can parameterize the location of the two ends points of this interval as
\be
(y_1,\bar{y}_1)=(\ep+it+i\sigma_1,\ep+it-i\sigma_1),\ \ \mbox{and} \ \ \ \
(y_2,\bar{y}_2)=(\ep+it+i\sigma_2,\ep+it-i\sigma_2). \label{valuey}
\ee

\begin{figure}
\begin{center}
\includegraphics[height=6cm,clip]{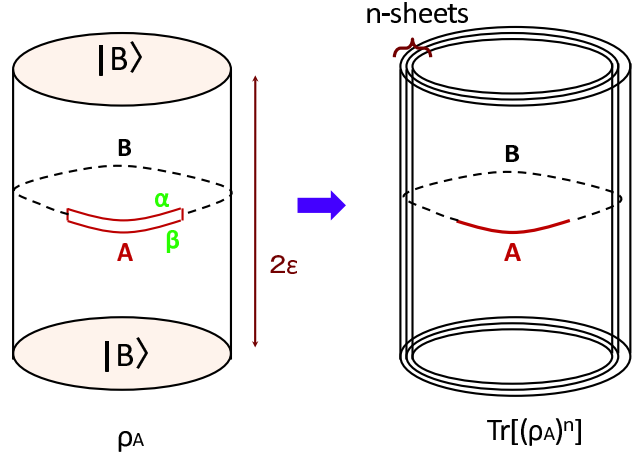}
\end{center}
\caption{The path-integral calculation of the reduced density matrix $[\rho_A]_{\ap\beta}$ (left) and
the trace Tr$\rho_A^n$ (right). }\label{PathF}
\end{figure}

 In the path-integral description \cite{CC,CCa},
the reduced density matrix $\rho_A(t)$ is obtained from $\rho_{tot}(t)$ (\ref{totd}) by gluing
the subsystem $B$ and the two boundaries on the subsystem $A$ are left as they are (see the left figure in
Fig.\ref{PathF}).
Then the important quantity $\mbox{Tr}[\rho_A(t)^N]$ can be found as the partition function of the free
Dirac fermion on a $N$-sheeted Riemann surface $\Sigma_N$ \cite{CC} as shown in the right figure in
Fig.\ref{PathF}. $\Sigma_N$ is defined by the N copies of
the cylinders (\ref{cyl}) which are glued successively along the subsystem A. This quantity is used to
define so called the Renyi entropy. Recently, remarkable properties of Renyi entropy and their 
consistency with holographic results have recently been studied in \cite{Matt}.

In this way, the entanglement entropy can be computed via
\be
S_A(t)=-\f{\de}{\de N}\log \mbox{Tr}[\rho_A(t)^N]|_{N=1}.  \label{dif}
\ee

By employing the replica trick, we can regard the system is given by
 $N$ free Dirac fermions $(\psi^{(a)},\bar{\psi}^{(a)})\ \ \ a=0,1,\ddd,N-1\in Z_{N}$
 on a single cylinder with the twisted boundary conditions
 at the two points $\psi^{(a)}_L(e^{2\pi i}y_1)=\psi^{(a+1)}_L(y_1)$ and
 $\psi^{(a)}_L(e^{2\pi i}y_2)=\psi^{(a-1)}_L(y_2)$.
Since the lagrangian is gaussian, its form remains unchanged even after we perform
the discrete Fourier transformation
$\psi^{(a)}\to \f{1}{\s{N}}\sum_{b=0}^{N-1}e^{\f{2\pi iab}{N}}\psi^{(b)}$ \cite{CaHu}. The interactions
between $N$ fermions only appear through the twisted boundary condition which are now
diagonalized as follows
\be
\psi^{(a)}_L(e^{2\pi i}y_1)=e^{\f{2\pi ia}{N}}\psi^{(a)}_L(y_1), \ \ \ \
\psi^{(b)}_L(e^{2\pi i}y_2)=e^{-\f{2\pi ia}{N}}\psi^{(b)}_L(y_2). \label{tbc}
\ee
These boundary conditions can be equally replaced with the twisted sector vertex operators
$\sigma^{(a)}$ with the lowest conformal dimensions. Finally, we can calculate the entanglement entropy
from the formula
\ba
&&\mbox{Tr}[\rho_A(t)^N]=\prod^{\f{N-1}{2}}_{a=-\f{N-1}{2}}\la \sigma^{(a)}(y_1,\bar{y}_1)\sigma^{(-a)}(y_2,\bar{y}_2)\lb_{cylinder} \no
&&\ \ \ \ \ =\prod^{\f{N-1}{2}}_{a=-\f{N-1}{2}}\f{\la B|e^{-2\ep H}\sigma^{(a)}(y_1,\bar{y}_1)\sigma^{(-a)}(y_2,\bar{y}_2)|B\lb}{\la B|e^{-2\ep H}|B\lb}. \label{defrhon}
\ea
Here we normalized the two point functions such that $\mbox{Tr}[\rho_A(t)]=1$.

\subsection{Bosonization and Two Point Functions on Cylinder}

A useful fact of the free massless Dirac fermion $(\psi,\bar{\psi})$
is that it can be bosonized into a free scalar $X$ via the standard formula
\be
\psi_L(y)=e^{iX_L(y)},\ \ \bar{\psi}_L(y)=e^{-iX_L(y)}, \ \ \psi_R(\bar{y})=e^{iX_R(\bar{y})},\ \
\bar{\psi}_R(\bar{y})=e^{-iX_R(\bar{y})}.
\ee
Our normalization of the two dimensional CFT
in this paper is $\al=2$ in string theory world-sheet \cite{Pol} and
the free Dirac fermion is equivalent to the compact boson $X$ at the radius $R=1$.
If we write the primary fields as $V_{(k_L,k_R)}(y,\bar{y})=e^{ik_LX_L(y)+ik_RX_R(\bar{y})}$,
then there are two candidates of the twisted vertex operators, which lead to the twisted
boundary conditions (\ref{tbc}),
\ba
&& \sigma^{(a)}_1(y,\bar{y})=V_{(\f{a}{N},-\f{a}{N})}(y,\bar{y})
=e^{i\f{a}{N}(X_L(y)-X_R(\bar{y}))},\label{twistb}\\
&& \sigma^{(a)}_2(y,\bar{y})=V_{(\f{a}{N},\f{a}{N})}(y,\bar{y})
=e^{i\f{a}{N}(X_L(y)+X_R(\bar{y}))},\label{twistc}
\ea
 which have the lowest dimensions if $a$ runs $a=-\f{N-1}{2},-\f{N-3}{2},\ddd,\f{N-1}{2}$ \cite{ANT}.

What we need to calculate to obtain the entanglement entropy are the two point functions of (\ref{twistb}) or (\ref{twistc}) on the cylinder. This can be found by performing calculations using the explicit form the boundary state $|B\lb$
(see e.g. \cite{Boundary}) as summarized in the appendix A. One may think that there are two choices i.e.
 the Neumann and Dirichlet boundary condition in the free scalar theory.
If we returns to the
 Dirac fermion theory, the Neunmann boundary condition relates $\psi_{L}$ and
 $\bar{\psi}_{L}$ to $\psi_R$ and $\bar{\psi}_{R}$ at the boundary, while the Dirichlet one does
 $\psi_{L,R}$ to $\bar{\psi}_{R,L}$, respectively. We can see that the discrete Fourier transformation,
employed to map the replicated fermions into the new ones which satisfy (\ref{tbc}), is consistent
only when we choose $\sigma^{(a)}_1$ or $\sigma^{(a)}_2$ for the Neumann or Dirichlet boundary condition,
respectively.

The final result of the normalized two point functions in both Neumann and Dirichlet is given by the
same expression
\be
\la \sigma^{(a)}(y_1,\bar{y}_1)\sigma^{(-a)}(y_2,\bar{y}_2)\lb
=\left(\f{\eta(\f{2i\ep}{\pi})^6\cdot|\theta_1(\f{\ep+it}{\pi i}+\f{\sigma}{2\pi}|\f{2i\ep}{\pi})|
|\theta_1(\f{\ep+it}{\pi i}-\f{\sigma}{2\pi}|\f{2i\ep}{\pi})|}
{|\theta_1(\f{\sigma}{2\pi}|\f{2i\ep}{\pi})|^2\cdot
|\theta_1(\f{\ep+it}{\pi i}|\f{2i\ep}{\pi})|^2}\right)^{\f{a^2}{N^2}}, \label{NB}
\ee
where we defined $\sigma\equiv\sigma_2-\sigma_1$. Refer to appendix A for a
derivation of this result (\ref{NB}). About the definition of the eta function $\eta(\tau)$
and theta function $\theta_1(\nu|\tau)$ we followed the convention in \cite{Pol}.

\subsection{Time Evolution of Entanglement Entropy with Finite Size}

By substituting the two point functions (\ref{NB})
into (\ref{defrhon}) and (\ref{dif}), we can calculate the entanglement entropy $S_A(t)$.
Using the formula
\be
\sum_{a=-\f{N-1}{2}}^{\f{N-1}{2}}\f{a^2}{N^2}=\f{1}{12}\left(N-\f{1}{N}\right),
\ee
we obtain the entanglement entropy
\be
S_A(t,\sigma)=\f{1}{6}\log \f{|\theta_1(\f{\sigma}{2\pi}|\f{2i\ep}{\pi})|^2\cdot
|\theta_1(\f{\ep+it}{\pi i}|\f{2i\ep}{\pi})|^2}{\eta(\f{2i\ep}{\pi})^6\cdot|\theta_1(\f{\ep+it}{\pi i}+\f{\sigma}{2\pi}|\f{2i\ep}{\pi})|
|\theta_1(\f{\ep+it}{\pi i}-\f{\sigma}{2\pi}|\f{2i\ep}{\pi})|\cdot a_{UV}^2}, \label{EEN}
\ee
where we recovered the cut off $a_{UV}$ dependence in an obvious way. Notice that this is independent
from the choice of the boundary condition (Neumann or Dirichlet one).
From this it is clear that $S_A(t)$ satisfies
\be
S_A(t,\sigma)=S_A(t,2\pi-\sigma)=S_B(t,\sigma). \label{purep}
\ee
This is consistent with the assumption that the total system is a pure state and the von-Neumann entropy for the total system is vanishing.

\begin{figure}
\begin{center}
\includegraphics[height=5cm,clip]{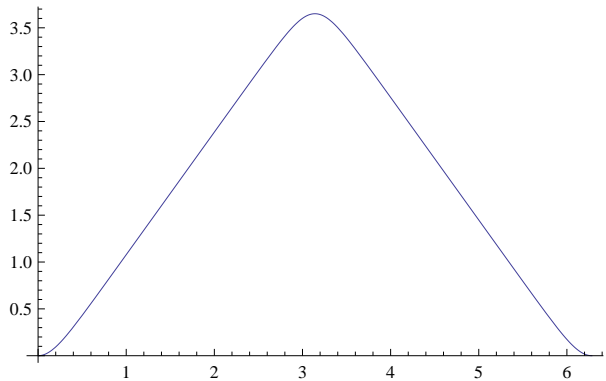}
\end{center}
\caption{The plot of $S_A(\pi/2,\sigma)-S_A(\pi/2,0)$ as a function of $\sigma$ at $\ep=0.2$.}\label{figEEo}
\end{figure}

\begin{figure}
\begin{center}
\includegraphics[height=5cm,clip]{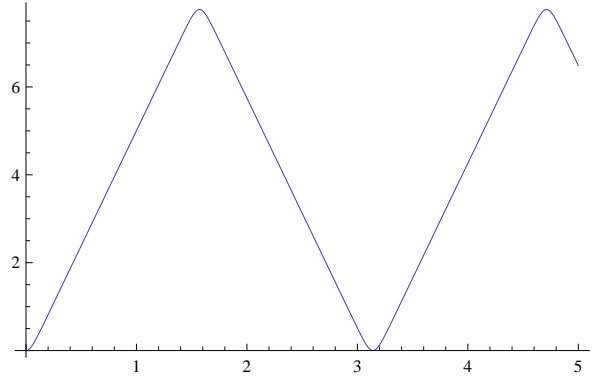}
\end{center}
\caption{The plot of $S_{eff}(t)\equiv 2\left(S_A(t,\pi)-S_A(0,\pi)\right)$ as a function of $t$
at $\ep=0.2$.}\label{figEEt}
\end{figure}

After the modular transformation, we can rewrite (\ref{EEN}) as follows\footnote{Actually, in this expression, the values of $\theta_1(\f{\ep+it}{2\ep}|\f{\pi i}{2\ep})$, $\theta_1(\f{\ep+it}{2\ep }+\f{i\sigma}{4\ep}|\f{\pi i}{2\ep})$ and $\theta_1(\f{\ep+it}{2\ep }-\f{i\sigma}{4\ep}|\f{\pi i}{2\ep})$ are always all real even before we take their absolute values. This justify our analytical continuation Re$[y]\to \ep+it$.}
\be
S_A(t,\sigma)=\f{1}{3}\log\f{2\ep}{\pi a_{UV}}+\f{1}{6}\log \f{|\theta_1(\f{i\sigma}{4\ep}|\f{\pi i}{2\ep})|^2\cdot
|\theta_1(\f{\ep+it}{2\ep}|\f{\pi i}{2\ep})|^2}{\eta(\f{\pi i}{2\ep})^6\cdot|\theta_1(\f{\ep+it}{2\ep }+\f{i\sigma}{4\ep}|\f{\pi i}{2\ep})||\theta_1(\f{\ep+it}{2\ep }-\f{i\sigma}{4\ep}|\f{\pi i}{2\ep})|}
\label{EENH}.
\ee
The explicit form of $S_A(t,\sigma)$ is plotted in Fig.\ref{figEEo} as a function of
$\sigma$ and Fig.\ref{figEEt} as a function of $t$.
In the limit $\ep\to \infty$, where the quench disappears, we reproduce the standard result
\cite{HLW,CC}
\be
S_A=\f{1}{3}\log \left(\f{2}{a_{UV}}\sin\f{\sigma}{2}\right).
\ee
In the opposite limit $\ep\to 0$, which corresponds to the infinitely extended space,
we can easily reproduce the known result in \cite{CCa} at the central charge $c=1$
\ba
S_A&=& S_{div}+\f{\pi t}{6\ep}\ \ \ (0<t<\f{\sigma}{2})\no
&=& S_{div}+\f{\pi \sigma}{12\ep}\ \ \ (t>\f{\sigma}{2}), \label{timeee}
\ea
where we separate the divergent part $S_{div}\equiv\f{1}{3}\log\f{2\ep}{\pi a_{UV}}$.
The physical meaning of the constant term $\f{\pi \sigma}{12\ep}$ can be qualitatively
understood as that of a thermal CFT gas with length $\sigma$ and at the effective temperature
\be
T_{eff}=\f{1}{4\ep}.  \label{teff}
\ee

We can also calculate the effective coarse-grained entropy introduced in (\ref{effE}) as follows
\be
S_{eff}(t)=2\left(S_A(t,\pi)-S_A(0,\pi)\right)=\f{2}{3}\log \f{\theta_2(0|\f{2i\ep}{\pi})\cdot
|\theta_1(\f{\ep+it}{\pi i}|\f{2i\ep}{\pi})|}{4\eta(\f{2i\ep}{\pi})^3\cdot|\theta_2(\f{\ep+it}{\pi i}|\f{2i\ep}{\pi})|},
\ee
which is clearly finite and does not depend on the cut off $a_{UV}$.
Then we notice a periodicity in time
\be
S_A(t,\sigma)=S_A(t+\pi,\sigma), \ \ \ \ S_{eff}(t)=S_{eff}(t+\pi), \label{entperi}
\ee
as is clear in in Fig.\ref{figEEt}.
This short periodicity is peculiar to our free CFT, where excitations triggered by the quench propagate
at the speed of light and return to exactly the same state after it goes half of the circle. This is because
the energy of oscillators included in the boundary state is always an even integer.
In generic field theories this does not occur. However, when the size of the space manifold is finite,
then the Poincare recurrence leads to a much longer periodicity of time
which is estimated as an exponential of a coarse-grained entropy $\Delta t\sim e^{S_{eff}}$, which has
been the crucial fact to resolve the information problem for eternal AdS black holes
\cite{MaE,FeLi,IiPo}.

\begin{figure}
\begin{center}
\includegraphics[height=5cm,clip]{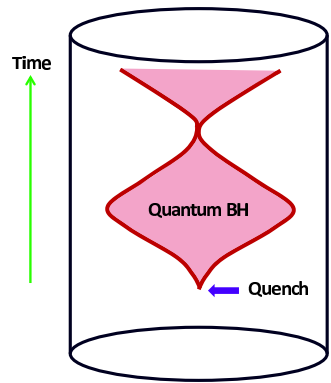}
\end{center}
\caption{Quantum black hole creations and annihilations in
AdS space by the quantum quench as obtained from Fig.\ref{figEEt}.}\label{BHcr}
\end{figure}

The AdS/CFT implies that a
holographic dual of gravity on AdS$_3$ with significant stringy corrections is
a weakly (or free) coupled field theory\footnote{Strictly speaking, in the standard examples of
AdS$_3/$CFT$_2$, we also need to take into account
the free scalar field sector in addition to the fermion sector we analyzed. However, since both are
free field theory, we expected that the qualitative result will not change.},
 which shows the short time periodicity like (\ref{entperi}).
In the gravity side, this oscillating behavior of $S_A(t)$ and $S_{eff}(t)$
can be interpreted as successive creations and
annihilations of black holes as sketched in Fig.\ref{BHcr}.
In this way, our free field theory example offers us a
holographic dual of black hole formations and evaporations in an extremely stringy region
and this is manifestly free from no information loss problem as the total system is described
by a pure state. Even though our argument on the information problem is along the line with
\cite{Page,MaE,FeLi,HaPr,SeSu,IiPo}, our approach is new in that we explicitly treat a real time-dependent
process of black hole formations by employing its holographic dual, rather than
focusing on the static black holes. Notice again that in this example, the size of black holes are probed
by the entanglement entropy and the standard total von-Neumann entropy is vanishing and not useful.

\section{Time-dependent Correlation Functions}

In previous section, we observed that $S_{eff}$ is oscillating in
our time-dependent system. We then identified the dynamics in CFT is
dual to successive productions and annihilations of quantum black
holes. We can naturally imagine the mechanism of its dynamics as
follows. First, a black hole is created by a quantum quench and
later it evaporates via a quantum analogue of Hawking radiations,
which are usually semiclassically analyzed. But the radiations are
completely reflected by the time-like boundary of the AdS. Therefore
they again concentrate at the center of AdS and recreate a black
hole. To confirm this idea, we would like to find time-dependent
signals of radiations.\footnote{For asymptotically flat black holes,
we can successfully calculate radiations from thermalized D-branes
by taking into account the coupling between open string fields and
close string fields as in \cite{DaMa}. Below we would like to
consider time-dependent radiations from black holes in AdS spacetime
rather than flat spacetime. } Though we will perform explicit
computations for a free Dirac fermion below, we can equally apply
the results in this section to correlation functions in a free
scalar theory compactified at $R=1$.

\subsection{One Point Functions}

Assume that an operator $O$ in CFT is dual to a field $\vp$ in AdS.
The bulk to boundary relation in AdS/CFT \cite{BB} argues that the
one point function $\la O(x) \lb$ in the CFT is proportional to the
value of the field $\vp$ near the AdS boundary. Since we are not
deforming the CFT itself, the profile of $\vp$ is supposed to be
normalizable.

The most important one point function in time-dependent backgrounds will be
that of the energy stress tensor.
Due to its conservation law, this clearly becomes time-independent.
It is straightforward to calculate the energy density ${\cal E}_n$ of the
left-moving oscillator $\ap_n$ from (\ref{totd}) as follows
\be
{\cal E}_n=\f{1}{2\pi}\la \ap_{-n}\ap_{n}\lb= \f{n}{2\pi(e^{4\ep n}-1)}. \label{onep}
\ee
This indeed obeys the Bose-Einstein distribution with the effective temperature estimated in
(\ref{teff}). This temperature can also be seen in the reduced density matrix when we trace out
the right-moving oscillators $\ti{\ap}_n$ (or equally left-moving ones)
\be
\rho_L=\mbox{Tr}_{R}e^{-\ep H}|B\lb \la B| e^{-\ep H}=
\prod_{m=0}^\infty\sum_{n=0}^\infty e^{-4\ep mn}
\f{(\ap_{-m})^n}{\s{n!}}|0\lb \la 0| \f{(\ap_{m})^n}{\s{n!}},\label{cano}
\ee
where we omit the zero modes. The total energy density in the $\ep\to 0$ limit can be found as usual
${\cal E}_{tot}\simeq 2\sum_{n=1}^\infty {\cal E}_n\simeq \f{\pi}{6}T_{eff}^2$.

\begin{figure}
\begin{center}
\includegraphics[height=5cm,clip]{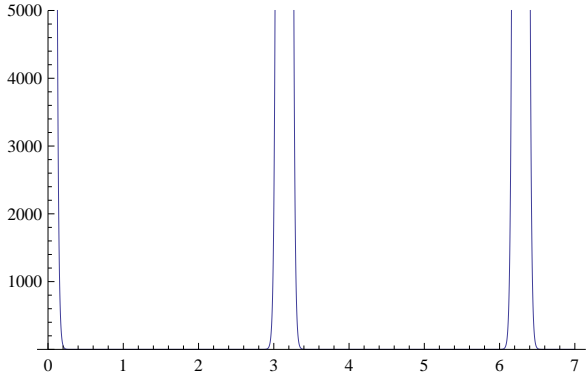}
\end{center}
\caption{The plot of the one point function
$e^{ikX(t,0)}$ as a function of $t$ at $\ep=0.2$ and $k=3$.}\label{oneplot}
\end{figure}

Even though we confirmed that our system is thermalized, we cannot
find signals of black hole creation and annihilation from the
time-independent one point function (\ref{onep}). Motivated by this
we would like to calculate another one point function which is not
dual to any conserved quantities. Especially we consider the one
point function  $\la e^{ikX(t,\sigma)}\lb$ for the Dirichlet
boundary state for the initial state $|\Psi_0\lb$, where $k$ can
take only integer values as we set $R=1$. After a calculation
similar to the appendix A, we finally obtain (we set $\sigma=0$) \be
\la e^{ikX(t,0)}\lb=\f{\theta_3\left(\f{k(\ep+it)}{\pi i}|\f{2\ep
i}{\pi}\right)} {\theta_3\left(0|\f{2\ep
i}{\pi}\right)}\cdot\f{|\eta\left(\f{2\ep i}{\pi}\right)|^{3k^2}}
{|\theta_1\left(\f{\ep+it}{\pi i}|\f{2\ep i}{\pi}\right)|^{k^2}}.
\ee As is clear from the plot in Fig.\ref{oneplot}, it has peaks at
$t=0,\pi,2\pi,\ddd$. This behavior is actually natural from our
interpretation in terms of black hole creations and annihilations.
As the result of the entanglement entropy suggests, a black hole is
created at $t=0$ and reaches its maximal size at $t=\f{\pi}{2}$.
Therefore, we expect that the radiations from the black hole will be
the most strong at $t=\f{\pi}{2},\f{3\pi}{2},\ddd$, remembering the
periodicity $\pi$. They will reach at the AdS boundary after the
propagation time\footnote{ This can be found from the metric of the
global AdS$_3$: $ds^2=-\cosh^2 \rho dt^2+d\rho^2+\sinh^2\rho
d\theta^2$ as $\Delta t=\int^\infty_0
\f{d\rho}{\cosh\rho}=\f{\pi}{2}$. In the presence of black holes one
may think that the propagation time of a massless particle will be
changed. Indeed, in a semiclassical black hole, this happens and
affects the structures of singularities of two point functions
\cite{HLR}. However, we do not consider this modification as we are
interested in a region in gravity where quantum corrections are so
large that the even horizon does not seem to be defined using a
metric.} $\Delta t=\pi/2$. This explains the peaks of $\la
e^{ikX(t,0)}\lb$ since the one point function is dual to the value
of bulk scalar field near the AdS boundary and its square should be
proportional to the strength of radiations. Notice that this
non-vanishing one point function is peculiar to time-dependent black
holes. This is because the one point function, which is
holographically dual to a classical radiation, is vanishing in the
static thermal CFT.

\subsection{Two Point Functions}

The another reason that we are interested in time dependent correlation functions is because we would like to discuss the information loss problem of black holes \cite{Ha}.
In \cite{MaE,FeLi,IiPo}, it has been argued that information loss problem is dual to the quasinormal behavior (exponential decay) of time dependent
two point functions in a large N gauge theory at finite temperature:
\begin{equation}
\la O(t_1)O(t_2)\lb\sim \exp [ -c|t_1-t_2| ].
\end{equation}
where $c$ is a positive real constant. These behavior breaks the unitarity of the theory because this shows that any fluctuations that
we add to the initial state will eventually vanish. In the same papers, it has also been conjectured in a finite $N$ gauge theory, the decay of the correlation stops at the value of order
$\exp[-O(N^2) ]$, then correlation grows, and repeats
this behavior because of the Poincare recurrence of the theory.

Indeed, this quasinormal behavior can be seen in our model.
If we set $y_1=\epsilon+it$, $y_2=\epsilon$ and
$k_L=k_R=k$ in the two point functions (\ref{eq:zetton}) for the Neumann boundary condition,
we find \begin{eqnarray}
G(t) &=&  \la \exp ik X(t,0) \exp  -ik X(0,0) \lb_N  \nonumber \\
&=&  \eta \left(\frac{2i\epsilon}{\pi} \right)^{6k^2}\cdot
\left|\frac {  \theta_1 \left(\frac{(\epsilon+it)}{\pi i} | \frac{2i\epsilon}{\pi}\right)  }
{\theta_1 \left(\frac{t}{2\pi } | \frac{2i\epsilon}{\pi}\right) ^2}  \right|^{k^2}\cdot
\left|\frac {  \theta_1 \left(\frac{\epsilon}{\pi i} | \frac{2i\epsilon}{\pi}\right)  }
{\theta_1 \left(\frac{2\epsilon+it}{2\pi i} | \frac{2i\epsilon}{\pi}\right)^2}  \right|^{k^2}.
\end{eqnarray}

\begin{figure}
\begin{center}
\includegraphics[height=5cm,clip]{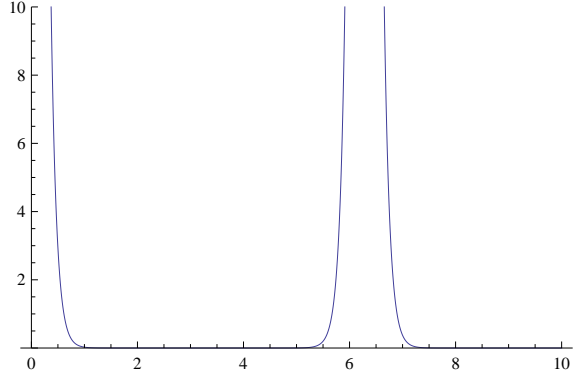}
\end{center}
\caption{The plot of $G(t)$ as a function of t at $\ep=0.2$.}\label{Gtplot}
\end{figure}

In Fig.\ref{Gtplot}, we show the plot of $G(t)$. It has the periodicity $2\pi$ as a signal
propagates from $\sigma=0$ and it has to go back to the same point. The divergences at $t\in 2\pi$Z
 come from the usual short distance behavior of the operator product. If $ \epsilon \ll t \ll 1$, we can
find the following exponential behavior of $G(t)$
\begin{eqnarray}
G(t)\simeq f(\ep)
 \exp \left[ -\frac{\pi k^2 t}{2\epsilon} \right],
\end{eqnarray}
for a certain time-independent function $f(\ep)$. Due to the $2\pi$ periodicity, after $G(t)$ gets decreased
exponentially, it reaches its (non-zero) minimum at $t=\pi$ and then again begins to increase.

Finally, it is also intriguing to study the time evolution of the two point function with separated points
$\la\exp  ik X(t,\sigma) \exp  - ik X(t,0)\lb$  for Neumann boundary condition.
If we use the formula (\ref{eq:zetton}) of Appendix A, and set $y_1=\epsilon+it+i\sigma$ and
$y_2=\epsilon+it$, we have
\begin{eqnarray}
F(t,\sigma)
&=&\la\exp  ik X(t,\sigma) \exp  - ik X(t,0)\lb_N \no
&=&
 \eta \left(\frac{2i\epsilon}{\pi} \right)^{6k^2}\cdot\frac{\left| \theta_1\left(\frac{\epsilon+it}{\pi i}|\frac{2i\epsilon}{\pi}\right)\right|^{2k^2}}
{\left|\theta_1\left(\frac{\sigma}{2\pi }|\frac{2i\epsilon}{\pi}\right)\right|^{2k^2}\left| \theta_1\left(\frac{2(\epsilon+it)+i\sigma}{2\pi i}|\frac{2i\epsilon}{\pi}\right)\right|^{k^2}
\left| \theta_1\left(\frac{2(\epsilon+it)-i\sigma}{2\pi i}|\frac{2i\epsilon}{\pi}\right)\right|^{k^2}}.\nonumber
\end{eqnarray}

\begin{figure}
\begin{center}
\includegraphics[height=5cm,clip]{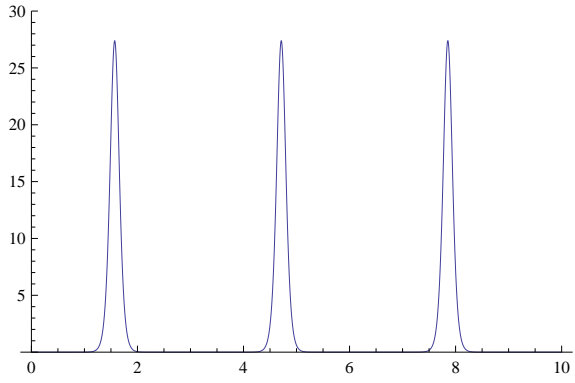}
\end{center}
\caption{The plot of $F(t,\sigma)$ as a function of $t$ at $\ep=0.2$.}\label{Ftplot}
\end{figure}

In Fig.\ref{Ftplot}, we plot $F(t,\sigma)$. The graph shows that the correlation becomes strong when $S_{eff}$ is large (compare with Fig.\ref{figEEt}). This can be regarded as a further evidence
of the successive productions and annihilations of quantum black holes in our system.

\section{Conclusions}

In the first part of this paper, we pointed out a puzzle. Because a
thermalization process in the CFT side should be unitary, the
von-Neumann entropy of the final state must remain vanishing. On the
other hand, in its gravity dual, we seem to have non-vanishing
entropy as the final state includes a black hole. These observations
look contradict with each other. However this is not the case. We
showed that even in the gravity side, the von-Neumann entropy of the
final state is zero by using the property $S_A=S_B$ of holographic
entanglement entropy for an arbitrary pure state. We proved this by
making use of the fact that the apparent horizon of the black hole
vanishes in the past. We then proposed a direct generalization of
thermal entropy by using the entanglement entropy. From a
holographic point of view, this quantity seems reasonable because $
S_{eff} $ is essentially proportional to the area of the apparent
horizon of the corresponding black hole.

Nevertheless, if quantum corrections are taken into account, one
might think that the entropy of the final state in the gravity dual
can be also nonzero due to the Hawking radiation. This observation
is closely related to the information loss problem. In the latter
part of this paper, we calculated the time evolution of entanglement
entropy following a quantum quench. Especially, we considered two
dimensional Dirac fermions on a cylinder. The result shows that the
von-Neumann entropy is always vanishing and the coarse-grained
entropy $S_{eff} $ is oscillating as a function of time. This offers
us a clear toy holographic dual of a formation and decay of black
holes, which is free from the information problem. We interpreted
this fact as the successive productions and annihilations of quantum
black holes. We also studied the behavior of one and two point
functions and confirmed that they support this interpretation. We
found that the two point functions cannot be zero and is periodic in
time which shows some recurrence of our free system.

\vskip6mm
\noindent
{\bf Acknowledgments}
We are grateful to M. Headrick, W. Li, S. Ryu, Y. Srivastava and S. Sugimoto for useful comments.
TT and TU are supported by World Premier International
Research Center Initiative (WPI Initiative), MEXT, Japan.
The work of TT is also supported in part by JSPS
Grant-in-Aid for Scientific Research No.20740132, and by JSPS
Grant-in-Aid for Creative Scientific Research No.\,19GS0219.  TT is
supported by World Premier International Research Center Initiative
(WPI Initiative), MEXT, Japan.

\newpage
\appendix
\section{A Derivation of Two Point Functions on Cylinder}
Here we would like to present an explicit derivation of two point functions (\ref{NB}) on cylinder.
We work in the description of a massless scalar field $X$ via the bosonization of the
Dirac fermion. We set $\al=2$ in string theory convention and follow \cite{Pol} about the notations
of the mode expansions of a scalar field and the definitions of theta functions.

We employ the complex coordinate $(y,\bar{y})$ on the cylinder defined by
(\ref{cyl}). In the operator formalism, the scalar field is quantized as usual
\ba
&& X_L=x_L-ip_Ly+i\sum_{m\neq 0}\f{\ap_m}{m}e^{-my},\no
&& X_R=x_R-ip_R\bar{y}+i\sum_{m\neq 0}\f{\ti{\ap}_m}{m}e^{-m\bar{y}},
\ea
where the commutation relations are given by
\be
[x_L,p_L]=i,\ \ \  [x_R,p_R]=i,\ \ \ [\ap_m,\ap_{n}]=m\delta_{n+m,0},\ \ \
[\ti{\ap}_m,\ti{\ap}_{n}]=m\delta_{n+m,0}.
\ee
At the compactification radius $R$, the momenta are quantized as follows
\be
p_L=\f{n}{R}+\f{wR}{2},\ \ \ \ \  p_R=\f{n}{R}-\f{wR}{2},
\ee
in terms of the integers $n$ (momentum) and $w$ (winding).
The (un-normalized) two point function of the normal ordered vertex operator
$V_{(k_L,k_R)}=:e^{ik_LX_L+ik_RX_R}:$ is written as follows
\ba
&& \la V_{(k_L,k_R)}(y_1,\bar{y}_1) V_{(-k_L,-k_R)}(y_2,\bar{y}_2)\lb_{cylinder} \no
&& \ =\la B|e^{-2\ep H}  V_{(k_L,k_R)}(y_1,\bar{y}_1) V_{(-k_L,-k_R)}(y_2,\bar{y}_2)|B\lb. \label{twoc}
\ea
The Hamiltonian is given in terms of the Virasoro generators as $H=L_0+\bar{L}_0-\f{1}{12}$.

\subsection{Neumann Case}

The boundary state $|B\lb$ for the Neumann boundary condition \cite{Boundary} is given by
\be
|B\lb_N={\cal N}~ e^{-\sum^\infty_{n=1}\f{1}{n}\ap_{-n}\ti{\ap}_{-n}}\sum^\infty_{w=-\infty}|w\lb,
\ee
where ${\cal N}$ is the normalization factor whose explicit value is not necessary for our
purpose. For bosonization procedures in the boundary state formalism, refer to \cite{Boson}.

The zero-mode part of (\ref{twoc}) can be calculated by using the Baker-Campell-Hausdorff (BCH) formula
as follows
\ba
&& \sum_{w=-\infty}^\infty \la w| e^{-2\ep H}e^{ik_L(x_L-ip_Ly_1)+ik_R(x_R-ip_R\bar{y}_1)}
e^{-ik_L(x_L-ip_Ly_2)-ik_R(x_R-ip_R\bar{y}_2)}|w\lb\no
&& =\sum_{w=-\infty}^\infty e^{-\f{R^2w^2\ep}{2}}e^{\f{R}{2}
\left(k_Lw(y_1-y_2)-k_Rw(\bar{y}_1-\bar{y}_2)\right)}
e^{\f{k_L^2}{2}(y_1-y_2)+\f{k_R^2}{2}(\bar{y}_2-\bar{y}_1)}.
\ea

The massive modes can be computed by employing the BCH formula, leading to the identity
\be
\la 0|e^{-\hat{\ap}\hat{\beta}z}e^{a_L\hat{\ap}+a_R\hat{\beta}}e^{b_L\hat{\ap}^++b_R\hat{\beta}^+}
e^{-\hat{\ap}^+\hat{\beta}^+}|0\lb=\f{1}{1-z}\cdot e^{\f{a_Lb_L+a_Rb_R-a_La_R-zb_Lb_R}{1-z}}, \label{forn}
\ee
where we assume the commutation relation
$[\hat{\ap},\hat{\ap}^+]=[\hat{\beta},\hat{\beta}^+]=1$.
Especially we can perform the calculations for massive parts by taking $z=e^{-4n\ep}$ and
\ba
&&  a_L=-\f{k_L}{\s{n}}(e^{-ny_1}-e^{-ny_2}),\ \ \ a_R=-\f{k_R}{\s{n}}(e^{-n\bar{y}_1}-e^{-n\bar{y}_2)}), \no
&& b_L=\f{k_L}{\s{n}}(e^{ny_1}-e^{ny_2}),\ \ \ b_R=\f{k_R}{\s{n}}(e^{n\bar{y}_1}-e^{n\bar{y}_2}),
\ea
in (\ref{forn}). By expanding $(1-e^{-4n\ep})^{-1}=\sum_{m=0}^\infty e^{-4mn\ep}$ and
changing the order of the summation w.r.t $n$ and $m$, eventually we can factorized the result into
$\theta$ functions. The final result is given by
\ba
&&\la B|e^{-2\ep H}  V_{(k_L,k_R)}(y_1,\bar{y}_1) V_{(-k_L,-k_R)}(y_2,\bar{y}_2)|B\lb_{N} \no
&& ={\cal N}^2\left[\sum_{w=-\infty}^\infty e^{-\f{R^2w^2\ep}{2}}e^{\f{R}{2}
\left(k_Lw(y_1-y_2)-k_Rw(\bar{y}_1-\bar{y}_2)\right)}\right]\cdot \f{1}{\eta\left(\f{2i\ep}{\pi}\right)}\no
&&\ \ \cdot\left(\f{\eta\left(\f{2i\ep}{\pi}\right)^3}{\theta_1\left(\f{y_2-y_1}{2\pi i}|\f{2i\ep}{\pi}\right)}\right)^{k_L^2}\cdot\left(\f{\eta\left(\f{2i\ep}{\pi}\right)^3}
{\theta_1\left(\f{\bar{y}_2-\bar{y}_1}{2\pi i}|\f{2i\ep}{\pi}\right)}\right)^{k_R^2}\cdot
\left(\f{\theta_1\left(\f{y_1+\bar{y}_1}{2\pi i}|\f{2i\ep}{\pi}\right)\theta_1\left(\f{y_2+\bar{y}_2}{2\pi i}|\f{2i\ep}{\pi}\right)}{\theta_1\left(\f{y_1+\bar{y}_2}{2\pi i}|\f{2i\ep}{\pi}\right)
\theta_1\left(\f{y_2+\bar{y}_1}{2\pi i}|\f{2i\ep}{\pi}\right)}\right)^{k_Lk_R}. \label{eq:zetton}
\ea
 After we substituting the values (\ref{valuey}) and
the twisted vertex operators (\ref{twistb}) at the free fermion radius $R=1$, we find that the
second line of the above expression is  rewritten as
\be
\f{\sum_{w=-\infty}^\infty e^{-\f{w^2\ep}{2}}}{\eta\left(\f{2i\ep}{\pi}\right)}
=\f{\theta_3\left(0|\f{2i\ep}{\pi}\right)+\theta_2\left(0|\f{2i\ep}{\pi}\right)}
{\eta\left(\f{2i\ep}{\pi}\right)},
\ee
where the first and second term are interpreted as the NS and R-sector of the Dirac fermion.
Therefore it is equal to
$\la B|e^{-2\ep H} |B\lb$ and can be neglected in order to find the normalized two
point functions as we did in (\ref{defrhon}). In this way we obtain the result (\ref{NB}).

\subsection{Dirichlet Case}

For the Dirichlet boundary condition, the boundary state is given by
\be
|B\lb_D={\cal N}'~ e^{\sum^\infty_{n=1}\f{1}{n}\ap_{-n}\ti{\ap}_{-n}}\sum^\infty_{n=-\infty}|n\lb.
\ee
By repeating similar calculations, in the end, we find the two point functions
\ba
&&\la B|e^{-2\ep H}  V_{(k_L,k_R)}(y_1,\bar{y}_1) V_{(-k_L,-k_R)}(y_2,\bar{y}_2)|B\lb_{D} \no
&& ={\cal N}^2\left[\sum_{n=-\infty}^\infty e^{-\f{2n^2\ep}{R^2}}e^{\f{n}{R}
\left(k_L(y_1-y_2)+k_R(\bar{y}_1-\bar{y}_2)\right)}\right]\cdot \f{1}{\eta\left(\f{2i\ep}{\pi}\right)}\no
&&\ \ \cdot\left(\f{\eta\left(\f{2i\ep}{\pi}\right)^3}{\theta_1\left(\f{y_2-y_1}{2\pi i}|\f{2i\ep}{\pi}\right)}\right)^{k_L^2}\cdot\left(\f{\eta\left(\f{2i\ep}{\pi}\right)^3}
{\theta_1\left(\f{\bar{y}_2-\bar{y}_1}{2\pi i}|\f{2i\ep}{\pi}\right)}\right)^{k_R^2}\cdot
\left(\f{\theta_1\left(\f{y_1+\bar{y}_1}{2\pi i}|\f{2i\ep}{\pi}\right)\theta_1\left(\f{y_2+\bar{y}_2}{2\pi i}|\f{2i\ep}{\pi}\right)}{\theta_1\left(\f{y_1+\bar{y}_2}{2\pi i}|\f{2i\ep}{\pi}\right)
\theta_1\left(\f{y_2+\bar{y}_1}{2\pi i}|\f{2i\ep}{\pi}\right)}\right)^{-k_Lk_R}. \nonumber
\ea
After we substituting the values (\ref{valuey}) and
the twisted vertex operators (\ref{twistc}) at the free fermion radius $R=1$, we find that the
second line of the above expression is rewritten as
\be
\f{\sum_{n=-\infty}^\infty e^{-2n^2 \ep}}{\eta\left(\f{2i\ep}{\pi}\right)}
=\f{\theta_3\left(0|\f{2i\ep}{\pi}\right)}
{\eta\left(\f{2i\ep}{\pi}\right)},
\ee
which includes only the NS-sector.

%%%%%%%%%% References %%%%%%%%%%%%%%%%%%%%%%%%%
\newcommand{\J}[4]{{\sl #1} {\bf #2} (#3) #4}
\newcommand{\andJ}[3]{{\bf #1} (#2) #3}
\newcommand{\AP}{Ann.\ Phys.\ (N.Y.)}
\newcommand{\MPL}{Mod.\ Phys.\ Lett.}
\newcommand{\NP}{Nucl.\ Phys.}
\newcommand{\PL}{Phys.\ Lett.}
\newcommand{\PR}{ Phys.\ Rev.}
\newcommand{\PRL}{Phys.\ Rev.\ Lett.}
\newcommand{\PTP}{Prog.\ Theor.\ Phys.}
\newcommand{\hep}[1]{{\tt hep-th/{#1}}}
%%%%%%%%%%%%%%%%%%%%%%%%%%%%%%%%%%%%%%%%%%%%%%%

\end{document}